\documentclass[12pt,preprint]{aastex}
\usepackage{natbib}
\usepackage{graphicx}
\shorttitle{Flat Circular Velocities on a megaparsec scale from the $\Lambda$CDM model}
\begin{document}
\title{Flat Circular Velocities on a megaparsec scale from the $\Lambda$CDM model}
\author{Antonino Del Popolo$^{1,2}$, Man Ho Chan$^3$}
\affil{$^1$ Dipartimento di Fisica e Astronomia, University of Catania, Viale Andrea Doria 6, 95123, Catania, Italy \\
$^2$ INFN Sezione di Catania, Via Santa Sofia 64, 95123 Catania, Italy\\
$^3$Department of Science and Environmental Studies, The Education University of Hong Kong, Tai Po, New Territories, Hong Kong, China}
\email{chanmh@eduhk.hk}

\begin{abstract}
A recent study using weak gravitational lensing reveals that there are some isolated galaxies having almost flat rotation curves at very large distance from the galactic centres. According to the authors of the study this provides a strong challenge the standard cold dark matter model, since the dark haloes are too small to explain their observations, especially for small stellar masses. In this article, we 
show that improving their model, the virial radius is larger than their estimates. The NFW rotational curve, and especially the pseudo Isothermal one, are in agreement with their flat rotational curves, especially for the larger baryonic mass bins used by the authors.    
\end{abstract}

\keywords{Dark Matter}

\section{Introduction}
Observational evidence shows that our universe contains a lot of invisible mass. One possible solution to the invisible mass problem is that there exists a large amount of unknown dark matter \citep{Bertone,Cirelli}. A large zoo of hypothetical particles like weakly interacting massive particles and axions are proposed to be the candidates of the cosmological dark matter \citep{Cirelli}. 

One of the most important evidence showing the existence of dark matter is the shape of galactic rotation curves. A rotation curve measures the function of the stellar speeds at different galactic distance $R$. Generally speaking, if there is no dark matter, the total mass distribution of a galaxy would be traced by ordinary matter. Since the density of ordinary matter usually follows an exponential distribution in $R$ \citep{Misiriotis}, one would expect the rotation curve of a galaxy showing a so-called Keplerian decline at large $R$. However, observational data indicate almost flat rotation curves appeared in most galaxies \citep{Sofue,Sofue1}. This suggests that a large amount of dark matter distributed throughout most of the galaxies. Moreover, the shape of a rotation curve could somewhat reveal the density distribution of dark matter in a galaxy.

Simulations show that if dark matter particles are collisionless, the dark matter density would follow the Navarro-Frenk-White (NFW) density profile \citep{Navarro1997}. This profile has been shown to give good agreements with the rotation curves in many galaxies, including our Milky Way galaxy \citep{Iocco} and the M31 galaxy \citep{Sofue2}. Since the NFW density profile goes like $\rho \propto R^{-3}$ at large $R$, we expect that most of the rotation curves at very large $R$ should slowly decline. High-quality data of our Milky Way galaxy collected by Gaia have shown a rotation curve decline at large $R$ \citep{Wang,Chan2023}. Therefore, at least in our Galaxy, this property (i.e. declining rotation curve at large $R$) is consistent with the prediction from the collisionless dark matter model. 

Recently, a new study, using weak gravitational lensing showed rotational curves of galaxies of different morphology
have flat rotation curves for very large radii, extending even beyond 1 Mpc \citep{Mistele}. According to the authors, this provides new evidence to support the prediction from Modified Newtonian Dynamics (MOND) that the galactic rotation curves should be flat in the deep-MOND regime (i.e acceleration is much less than the universal acceleration scale $a_0$). 
At the same time, they show that for the benchmark cold dark matter model, the galactic rotation curves decline beyond the dark matter virial radii.
From their Fig. 2, the observed rotation curves data generally do not agree with the cold dark matter model's prediction \citep{Mistele}. Therefore, this is a strong piece of evidence to support MOND and challenge the standard cold dark matter model. 

In this article, we critically examine this claim and show that it is enough to use a different abundance matching relation, more commonly used than that used by the authors of the paper, i.e \citep{Kravtsov2018}, obtain from that the $M_{100}$, and $R_{100}$, mass and virial radius to show that especially for the largest baryonic bin in Fig. 2 of \citet{Mistele}, the virial radius is larger than those calculated by the authors with the consequence that for the NFW rotational curve, and especially for the pseudo Isothermal one, we may obtain rotational curves in agreement with those of \citet{Mistele}. 
The paper is organized as follows: in Sect. 2 we discuss how the \citet{Mistele} rotational curves where compared the prediction of the NFW density profile, and how using \citet{Moster2013} abundance matching relation, and the typical overdensity for virialization in a $\Lambda$CDM model \citet{Mistele} result is modified. Sect. 3 is devoted to discussions. 

\section{Rotational curves in $\Lambda$CDM}      

In this section, we are going to show how some small changes to calculation of \citep{Mistele}, in particular in relation to the comparison of their rotational curves with the predictions of $\Lambda$CDM, as shown in their Fig. 2, is noteworthy modified doing the calculation in a more coherent way. In their paper, \citet{Mistele} inspired by \citet{Brower2021} analyzed a sample of isolated galaxies taken from the KiDS survey. Following the procedure in \citet{Mistele2024} they selected lens and source galaxy samples. The isolation criterion discussed in \citet{Mistele2024}, and \citet{Brower2021} is based on the KiDS photometric redshifts, which have significant uncertainties. At radii larger than $300 h_{70}^{-1} \rm kpc$, the isolated lens selection is affected by the photometric KiDS redshifts. By means of $\Lambda$CDM simulations, \citet{Brower2021} estimated that their weaker isolation criterion is reliable out to about 300 kpc. Studies in \citet{Mistele,Mistele2024} found empirically that the isolation criterion could be extended to $\simeq 1~ \rm Mpc$. So in \citet{Mistele}, they considered radii out to 1 Mpc, but they left the 300 kpc limit as a conservative lower bound. The rotation curve of a galaxy is obtained known the potential $\phi$ by
\begin{equation}
V_c=\sqrt{R|\frac{d\phi}{dR}|}.
\end{equation}
The potential, and the related rotational curve has been obtained historically using long slit spectra, and emission lines such as $H_\alpha$, and for a limited number of nearby galaxies, using the velocities of individual HII regions in galactic disks \citep{RubinFord1970}. Partly because its radial extent is greater (3, 4 times) than that of visible light, the HI line is a powerful tool to obtain rotational curves \citep{Bosma1981a,Bosma1981b}. Even with HI, one can study the rotational curves out to some tens of kpc. Using lensing, as shown by \citet{Brower2021,Mistele} one can reach much larger distances. As reported, around 300 kpc, as shown by \citet{Brower2021}, and $\simeq \rm 1~ Mpc$ according to \citet{Mistele}. In order to do this, one has to relate the lensing shear proﬁle to the physical excess surface density (ESD, denoted $\Delta \Sigma$) proﬁle through the critical surface density $\Sigma_{\rm crit}$
\begin{equation}
\Delta \Sigma (R) =\Sigma_{\rm crit} \gamma_t(R)= <\Sigma> (<R)-\Sigma(R),
\end{equation}
namely the average surface density $<\Sigma>(<R)$ within $R$ subtracted from the surface density $\Sigma(R)$ at projected radius $R$. Then the ESD can be converted into the circular velocity by means of
\begin{equation}
V_c(r)=\sqrt[•]{\frac{G (4 \Delta \Sigma(r) r^2)}{r}}=\sqrt[•]{4G \Delta \Sigma(r) r }.
\end{equation}
In their Fig. 2, \citet{Mistele} plotted the rotational curve obtained by lensing, and compared it with that 
predicted by a DM halo and a baryonic point mass $V^2_c(R)=V^2_{c,DM}(R)+GM_{b,eff}/R$ with averaged baryonic mass $M_{b,eff}$. For the DM rotational curve they choose the NFW profile. The halo parameters are obtained using \citet{Maccio2008} WMAP5 concentration parameter. For the four subplots in Fig. 2, they used the reported baryonic mass, and a stellar mass, not reported in the paper and that they gently gave to us. In order to get a rotational curve, if you use a NFW profile, you need the virial radius and the concentration parameter (obtained using \citet{Maccio2008}). Concerning the virial radius, they used the  Stellar-to-Halo Mass Relation (STHMR) obtained through abundance matching techniques, pionered by \citet{Moster2010}, and refined by \citet{Kravtsov2018,Moster2013,Behroozi2013}. The STHMR is a statistical method to determine the relationship between the stellar masses of galaxies and the masses of their host dark matter haloes. Then using a STHMR from the stellar mass one can obtain the virial mass. In Fig. 17 of \citet{Kravtsov2018} is shown a comparison between the STHMR of \citet{Kravtsov2018} with those of \citet{Moster2013,Behroozi2013}. The plot shows how for stellar masses larger than $2 \times 10^{10} M_\odot$ there is a large discrepancy between the \citet{Moster2013,Behroozi2013} relation and that of \citet{Kravtsov2018}. This last predicts much smaller virial radii with respect to the other two STHMR. The reason why the authors chose \citet{Kravtsov2018} STHMR is because according to them the last is more appropriate than \citet{Moster2013,Behroozi2013} `because, unlike other common relations, it does not overshoot the circular velocities at small radii in the high-mass bins'. However, with this choice, for example for the top subplot in Fig. 2, the stellar mass of $1.25 \times 10^{11} M_\odot$ gives a $M_{200}$ $\simeq 331 ~\rm kpc$, while the \citet{Moster2013} STHMR gives $\simeq 546 ~\rm kpc$, a big difference. 

According to early theoretical work \citep{Peebles1980}, a sphere of material will collapse if its density exceeds $1.686(1 +z)$ times that of the background. After the collapse there is the virialization phase, and the sphere will get the average density $\Delta \simeq 18 \pi^2 \simeq 178$ times the critical density. In the well known paper by \citet{Navarro1996}, approximated the value with $\Delta=200$, commonly used by that time to allow for easy comparison between papers. The value $\Delta \simeq 178$ is typical of an Einstein de-Sitter cosmology, but in the $\Lambda$CDM model the value is $\Delta \simeq 100$. For a flat Universe($\Omega_m+\Omega_{\Lambda}=1$), \citet{BryanNorman1998} gives $\Delta \simeq 18 \pi^2 -82\Omega_{\Lambda}-39 \Omega^2_{\Lambda}$, or see Fig. 2 of \citet{DelPopolo2013}. Then we have to convert the values of $M_{200}$ given by \citet{Kravtsov2018} to $M_{100}$. In order to do this conversion, we may use different ways, and we used the Eq. 8 in \citet{Lukic2009}
\begin{equation}
M_c = A(c) \left[ \ln \left( 1 + \sqrt[3]{\frac{200}{\Delta}} M_c c \right)
            - \frac{\sqrt[3]{\frac{200}{\Delta}} M_c c}
              {1+\sqrt[3]{\frac{200}{\Delta}} M_c c} \right]\ ,
\label{nfw_transform}
\end{equation}
where $M_c =M_{\Delta}/M_{200}$ and 
\begin{equation}
A(c) = \frac{1}{\ln (1+c) - c/(1+c)}\ .
\end{equation}
We have also to convert $c=c_{200}$ in $c_{100}$, 
as shown in \citet{Coe2010}. Finally, we get $M_{100}$ and $R_{100}$

%
%

\begin{table*}
\small
\caption{$M_{200}$, $R_{200}$, $M_{100}$, and $R_{100}$ for the Kravtsov and Moster STHMR. K stands for Kravtsov, M for Moster.}
 \label{table1}
 \begin{tabular}{@{}lcccccc}
  \hline
$M_{*} (M_\odot)$ & $M_{200,K} (M_\odot)$  &   $R_{200,K} (\rm kpc)$ & $M_{200,M} (M_\odot)$  &   $R_{200,M} (\rm kpc)$  & $M_{100,M} (M_\odot)$  &   $R_{100,M} (\rm kpc)$\\
  \hline
$1.25 \times 10^{11}$ & $4.19 \times 10^{12}$ &  330.55 &   $1.89\times 10^{13}$ &   545.99  &   $2.80 \times 10^{13}$  &  784.84\\
  \hline
$6.66 \times 10^{10}$  &  $1.62 \times 10^{12}$ & 240.80 &  $3.88 \times 10^{12}$ &  322.35 &  $5.49 \times 10^{12}$  &     455.92\\
\hline
$3.46 \times 10^{10}$ & $7.69 \times 10^{11}$ & 187.90 &  $1.01 \times 10^{12}$ &   206.09  &   $1.39 \times 10^{12}$ &      288.13 \\
\hline
$8.47 \times 10^9$  & $2.87 \times 10^{11}$ & 135.31 & $2.85 \times 10^{11}$ & 134.97 &  $3.79 \times 10^{11}$   &    186.93  \\
  \hline
 \end{tabular}
\end{table*}

The rotational curve obtained from a NFW density profile
\begin{equation}
\rho(R)=\frac{\rho_s}{\frac{R}{R_s}(1+\frac{R}{R_s})^2}
\end{equation}
is given by 
\begin{equation}
V^2(R)=\frac{GM_v}{R_v} \frac{1}{y} \frac{\log(1+cy)-\frac{cy}{1+cy}}{\log(1+c)-\frac{c}{1+c}}
\end{equation}
where $M_v$, and $R_v$ are the virial mass and radius, $c$ is the concentration parameter, and $y=R/R_v$. In \citet{Mistele}, $R_v=R_{200}$, $M_v=M_{200}$, and $c=c_{200}$. In this paper, we will also compare the rotational curves of \citet{Mistele}, with another theoretical rotational curve, obtained from the pseudo-Isothermal profile
\begin{equation}
\rho_ {\rm pISO}= \frac{\rho_0}{1+(\frac{R}{R_c})^2}  
\end{equation}
with a corresponding rotational curve
\begin{equation}
V(R)=\sqrt[•]{4 \pi G \rho_0 R^2_c [1-\frac{R_c}{R} \arctan(\frac{R}{R_c})]}
\end{equation}


The parameter $R_c$, and $R_s$ are different. A relation between the two parameters can be obtained following \citet{Boyarsky2009}. We may compare the NFW and pISO rotational curves with the data in Table 1. The main point of \citet{Mistele} paper is that the $\Lambda$CDM does not obtain curves flat in the same radial range. Then we normalize our rotational curves in order we may compare to \citet{Mistele} Fig. 2. 

The results of the comparison are shown in Fig. 1, and Fig. 2.

\begin{figure}
\vskip 10mm
 \includegraphics[width=150mm]{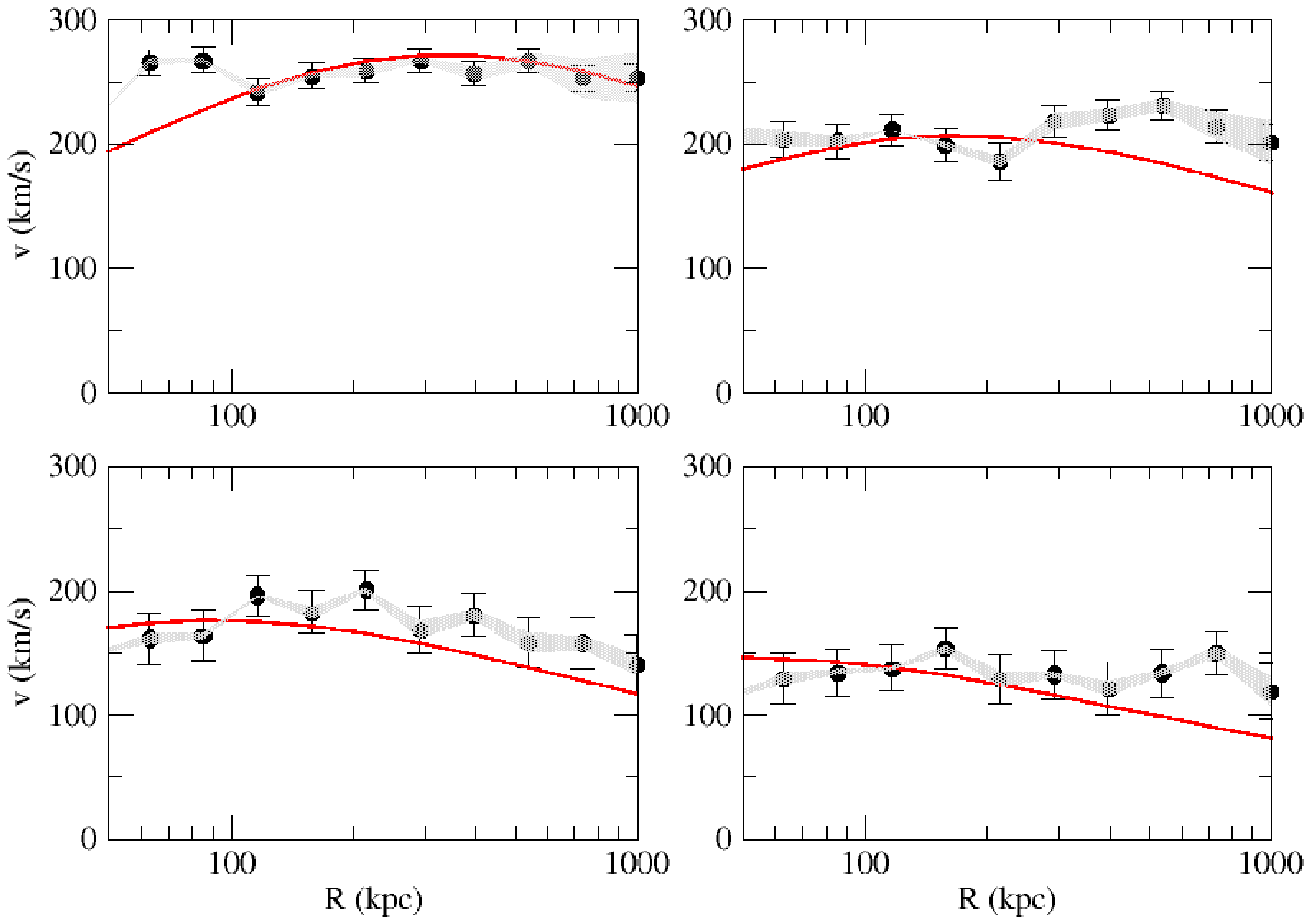}
 \caption{Comparison of the rotational curves in Fig. 2 of \citet{Mistele} with the NFW rotational curves with the described virial radius and concentration. The error bars and shaded band indicate the statistical errors and systematic errors respectively.}
\vskip 10mm
\end{figure}

\begin{figure}
\vskip 10mm
 \includegraphics[width=150mm]{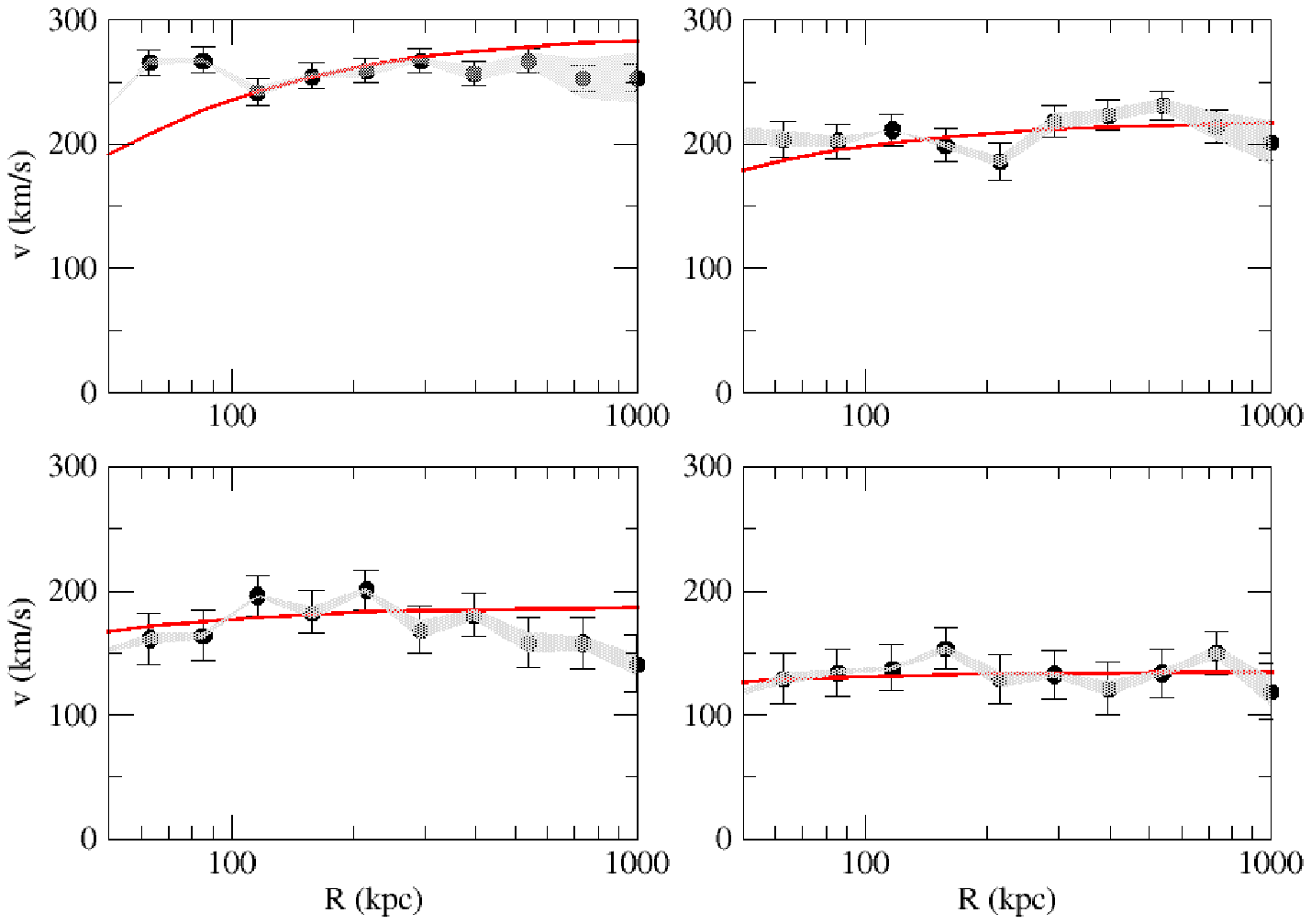}
 \caption{Comparison of the rotational curves in Fig. 2 of \citet{Mistele} with the pISO rotational curves with the described virial radius and concentration. The error bars and shaded band indicate the statistical errors and systematic errors respectively.}
\vskip 10mm
\end{figure}

\section{Discussion}

In the work of \citet{Mistele}, usual rotational curves obtained by gas kinematics are extended by more than an order of magnitude in distance from the galaxy center, and according to their result they remain flat to $\simeq  1$ Mpc. The authors even claim that beyond 1 Mpc there is not a clear departure from flatness, but at those radii the systematic uncertainties on  $V_c$ are large, and it is surely dubious that lenses are isolated. In this paper, we will not discuss if \citet{Mistele} results are really correct out to 1 Mpc, or out to 300 kpc as reported by \citet{Brower2021}. Future studies will discuss this issue. However, we want to recall that \citet{Brower2021}'s paper is strictly related to that of \citet{Mistele}. This last paper was stimulated by \citet{Brower2021} idea. In this paper, the authors used weak lensing to get the Radial Acceleration Realation (RAR), and at the same time, similarly to \citet{Mistele}, they obtained the rotation curves from the data of the fourth KIDS release. However, different from \citet{Mistele} the rotational curves were plotted till 300 kpc, which according to \citet{Brower2021} is the distance at which the uncertainty in photometric KIDS redshift can affect the the isolated lens selection. 

\citet{Mistele} shows that the rotational curves obtained from the NFW profile tend to fall at distances $R<1$ Mpc. The $V_c$ obtained through lensing seems to be flat out to $\simeq 1~ \rm Mpc$ even if there is an up and down behavior of the data. In one of the cases the rotation curve data seem to decline from 200 kpc. Following the discussion of the authors, there is not a large discrepancy between the NFW rotational curve and their data for radii smaller than 300 kpc, where as stressed by the authors the isolation criterion is sure. As reported, we are not interested in the correct range on which their analysis is correct but affirmations like that, let one think that they are not sure about the analysis to 1 Mpc. 
%
%

Since no galaxy is completely isolated, one should somehow take account of this fact. In particular the signal from correlated halos, the so called "two-halo" term, could somehow take account the previous issue. \citet{Mistele} tried to calculate this effect, and plotted in Fig. 2 the two-halo term, that somehow shows that this term decrease the discrepancy between their NFW fits and their data. However, the two-halo term was estimated by them just qualitatively, not quantitatively. Apart from the contributions to the isolated galaxies from nearby halos or higher redshift halos, a larger problem comes from  the uncertainty in the photometric KiDS redshifts which can affect the isolated lens selection. \citet{Brower2021} discussed this issue showing that it appears at radii larger than 300 kpc.

Concerning the discrepancy between the NFW rotational curve and those of the authors even at distances $< 300$ kpc, \citet{Mistele} claim that this discrepancy does not depend on the DM halo profile chosen, the STHMR chosen, or the mass-concentration relation. In this study, we verified if the last claims are real. As previously reported, we considered the STHMR \citep{Moster2013}, and the fact that in a $\Lambda$CDM Universe, virialization happens at $\Delta \simeq 100$.   
As shown in Table 1, there is a certain discrepancy between the virial radii obtained with the Kravtsov's or Moster's STHMR. The discrepancy increases if we use the typical $\Lambda$CDM $\Delta=100$ virial overdensity. For the first baryonic mass bin, the value of the virial radius is $R_{200}=330.55$ kpc for the Kravtsov's STHMR, and $R_{100}=784.84$ kpc for the Moster's STHMR, a factor $\simeq 2.4$ larger than the virial radius used by \citet{Mistele}. The difference decreases for smaller baryonic bin. The larger virial radii obtained in our estimate implies larger haloes, and we expect a better agreement between the \citet{Mistele} rotation curves and the NFW ones. In Fig. 1, we show the quoted comparison. The subplots, in clockwise direction corresponds to the subplots from top to bottom of \citet{Mistele} Fig. 2. The dots with error-bars are the estimated rotational curves by \citet{Mistele}, while the line gives the NFW rotational curve with the parameters in Table 1, normalized to match \citet{Mistele} velocity in their rotational curves. The top-left subplot shows a good agreement with the corresponding rotational curve in \citet{Mistele} . The first two points under 100 kpc are not of our interest, since we are interested in the behavior of the rotational curve at large radii. The top-right panel shows that the NFW rotational curve is in agreement with \citet{Mistele} till 300 kpc. The bottom-left rotational curve is in agreement with the data in \citet{Mistele}. The last bottom-right panel shows an agreement of the NFW rotational curve with that of \citet{Mistele} till 400 kpc. The disagreement shown on Fig. 2 of \citet{Mistele} is eliminated or noteworthy reduced. Then, we thought to use the rotational curve coming from the pseudo isothermal profile, and compared them to the data of \citet{Mistele} in our Fig. 2, as done with the NFW rotational curve. The plot shows that the $\Lambda$CDM rotational curves are flat at least till 1 Mpc. In conclusion, the large discrepancy between \citet{Mistele} rotational curves, depends from the STHMR used, the larger value of the virialization overdensity, and the density profile used for the fit (we refers to the pISO profile).

Before concluding, we want to stress that usually in literature, the fit of the rotational curves using the NFW profile is based on the usual $R_{200}$ virial radius which is not the correct one for the $\Lambda$CDM model, and in papers in which one has to convert from the stellar mass to the virial one, the STHMR several times is not chosen with prudence. This implies errors in the fits of the rotational curves in the papers published.

\section{Acknowledgements}
The work described in this paper was partially supported by the Dean's Research Fund from The Education University of Hong Kong (0400W) and a grant from the Research Grants Council of the Hong Kong Special Administrative Region, China (Project No. EdUHK 18300324).

\end{document}